# Semiconductor sub-micro-/nanochannel networks by deterministic layer wrinkling**

*Yongfeng Mei,\* Dominic J. Thurmer, Francesca Cavallo, Suwit Kiravittaya, Oliver G. Schmidt*

The wrinkling of thin films on substrate surfaces is a well-known phenomenon and has been studied in great detail with different material systems for several decades.[1-11] While a few potential applications of wrinkles have been put forward, such as force spectroscopy in cells,[1] optical devices,[3] metrology methods,[9] and flexible electronics,[11] it seems an intriguing and almost obvious idea to use wrinkles as complex nanochannel networks on a substrate surface to study nanofluidics[12-14] or to herald applications in bionanotechnology.[15-17] Recently, it has been suggested to employ folded thin films[18] with wrinkles running perpendicular to the main fold to realize complex nanochannel systems,[4] but fluid flow through such wrinkles has not been reported, so far. Here, we describe a technology that exploits the deterministic wrinkling and a subsequent bond-back of a semiconductor layer to create well-defined and versatile nanochannel networks. The technology is termed "Release and bond-back of layers (REBOLA)", and consists of the partial release, wrinkling, and bond back of a compressively strained functional layer on a substrate surface. Linear and circular nanochannel networks – both of which consist of a main channel and several perpendicularly oriented branch channels – are fabricated by REBOLA. In these networks, the periodicity and the positions of the branch channels can be tuned and controlled by changing the width of the partially released layers and by applying appropriate lithography. To elucidate the usefulness of REBOLA, we demonstrate nanofluidic transport as well as femto-litre filling and emptying of individual wrinkles on a standard semiconductor substrate.

The general concept of REBOLA is schematically outlined in Fig. 1a.[19] A thin strained functional layer, deposited on a sacrificial buffer layer is partially released from the substrate surface by selectively etching away the sacrificial buffer layer from one side. Once the strained functional layer is freed from the substrate, the strain elastically relaxes and causes wrinkles to form perpendicular to the etching front.[2a,20] Afterwards, the wrinkled functional layer bonds back to the substrate surface and forms a branched nanochannel network consisting of the self-formed wrinkles, which connect to the main channel running along the edge of the back-etched buffer layer. While Fig. 1a represents a generic description of REBOLA, which can be applied to many different material systems, geometries and layer thicknesses, we concentrate in this Letter on semiconductor based materials which are well-compatible to advanced Si and CMOS (complementary metal oxide semiconductor) technology. More precisely, we expose the surface of a SiGe on $SiO_2$ on Si (001) substrate to hydrofluoric acid. The acid gains access to the below lying $SiO_2$ through lithographically defined trenches, fine scratches or defects in the SiGe layer, and then etches away the insulator material a certain lateral distance ($L_e$). The compressively strained SiGe functional layer becomes detached form the substrate, wrinkles and eventually forms into a branched nanochannel network by partially bonding back to the surface. A schematic 3D illustration of the final structure is given as an inset in Fig. 1b.

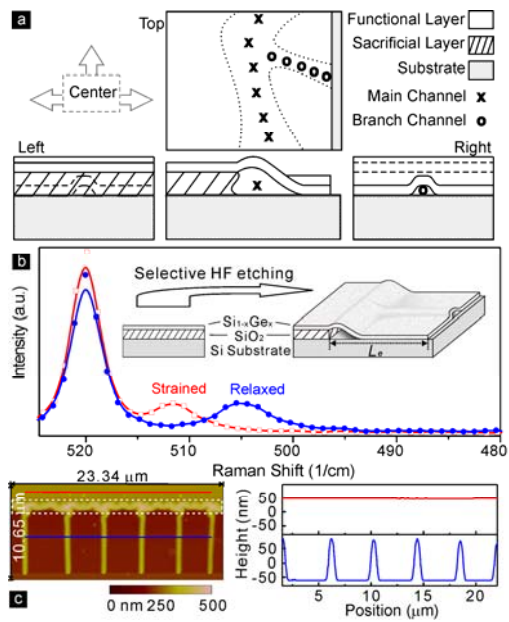

Figure 1: Fabrication and characterization of nanochannel networks by REBOLA. a, Generic description of the proposed micro-/nanochannel network. A single unit of such a network consists of a main channel and a perpendicularly oriented branch channel. b, Raman spectra of strained and relaxed ultra-thin $Si_{1-x}Ge_x$ film on 100 nm thick $SiO_2$ layer. The inset shows a schematic illustration of the formation of the $Si_{1-x}Ge_x$ channel structure by HF etching of $SiO_2$. c, AFM image of the linear network (left) and height profiles (right) along the lines indicated in the AFM image.

We measure the change in strain in the $Si_{1-x}Ge_x$ layer before and after REBOLA by micro-Raman scattering as presented in Fig. 1b. The red curve represents the Raman spectrum of the strained SiGe layer before underetching. The line at 520 1/cm stems from the first-order longitudinal optical phonon Si-Si vibration mode of the Si substrate, whereas the peak at 512 1/cm originates from the Si-Si vibration mode of the compressively strained SiGe layer. The blue spectrum is taken from an area where the SiGe layer was released and bonded back on the Si substrate surface. The Si-Si vibration mode of this SiGe layer is shifted by 7 1/cm compared to the strained SiGe layer. Using the equation[21] $\omega_{Si} = 520 - 62x + \Delta_{Si}\Sigma$, we determine a 24% average Ge composition and a 1.1% compressive strain in the $Si_{0.76}Ge_{0.24}$ film before etching.[22] The degree of relaxation can also be determined by measuring the increase in length of the wrinkled layer. An atomic force microscopy (AFM) image (left) and two linescans (right) are presented in Fig. 1c.: the red linescan was performed on the planar non-underetched surface and the blue one on the wrinkled area. The integrated path (20.717 µm) across the wrinkled area is 1.187% longer than the length of the linescan across the flat surface area (20.474 µm). Hence, the strain measurements by micro-Raman spectroscopy and the linescan method agree reasonably well. The small difference may be attributed to the inhomogeneous strain across the thin SiGe film in vertical direction and/or the not perfectly equidistant positions of the branch channels.

[*] Dr. Yongfeng Mei, Dominic J. Thurmer, Francesca Cavallo, Dr. Suwit Kiravittaya, Dr. Oliver G. Schmidt
Max-Planck-Institut für Festkörperforschung,
Heisenbergstrasse 1, D-70569 Stuttgart, Germany
E-mail: y.mei@fkf.mpg.de
Dr. Yongfeng Mei
Max-Planck-Institut für Festkörperforschung,
Heisenbergstrasse 1, D-70569 Stuttgart, Germany

[**] We thank Dr. S. Mendach, Dr. Ch. Deneke, Dr. M. Stoffel, and Dr. A. Rastelli for helpful discussions as well as U. Waizmann, A. Schulz, E. Coric, W. Winter, U. Zschieschang, and M. Riek for experimental assistance. We also thank Dr. Thamm at Karl Zeiss AG for the use of the Axiocam HSm. This work is financially supported by the BMBF (03N8711).

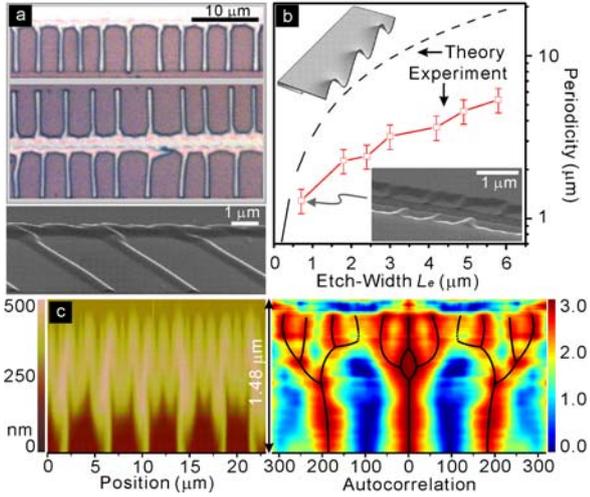

Figure 2: Experimental and theoretical analysis of linear nanochannel networks. a, Optical microscopy images of a linear nanochannel network with single-sided (upper) and double-sided branch channels (middle). An SEM image of a single-sided linear nanochannel network is given in the lower panel. b, Periodicity of wrinkles as a function of etch width. Dashed curve shows theoretical calculation for entirely free-hanging wrinkled film (upper left inset). The lower right inset shows a SEM image of a nanochannel network, in which the height is about 40 nm and the width is about 150 nm. c, AFM image (left) of the linear network near the etching front (marked by the dashed white rectangle in Fig. 1c) and its autocorrelation pattern (right).

REBOLA allows us to use standard lithography or simple mechanical scratching to define straight starting lines for the underetching. Two examples of linear nanochannel networks are shown in Fig. 2a. The network shown in the upper image of Fig. 2a consists of a single-sided branched channel network directly connected to the main channel running along the back-etched buffer layer. The middle image is a double-sided branched linear nanochannel network with the main channel running in between the wrinkled branch channels.

A scanning electron microscopy (SEM) image at the bottom of Fig. 2a reveals that the end of each branch channel, located at the starting edge, is open and that the opening has an arc shape with a width of 300-500 nm and a height of about 120 nm (See also Supplementary Fig. S1). This SEM confirms that the opposite branch channel ends are directly connected to the main channel running along the back-etched buffer layer. The average periodicity ($\lambda_0$) of the wrinkling (or the inter-distance between two branch channels) parallel to the etching front is about 3.3 µm with a standard deviation of 17.1%. As shown in Fig. 2b, $\lambda_0$ increases with $L_e$ in the experiment. A simple theoretical calculation, based on the minimization of elastic energy of a free-hanging wrinkled film (upper left inset of Fig. 2b),[23] is used to fit the experimental result. We recognize that the periodicities predicted by theory show the same trend as in experiment, but the values are 2-3 times larger. This discrepancy is attributed to the interaction of the free-hanging film with the substrate once the wrinkling amplitude becomes larger than the sacrificial layer thickness. If the wrinkled film begins to partially bond back to the substrate surface, the layer cannot adapt its equilibrium periodicity as calculated for an entirely free hanging film, and $\lambda_0$ remains smaller than predicted by theory with increasing $L_e$. However, the film is not expected to tightly bond back to the surface during underetching,[24] which means that $\lambda_0$ could still slightly increase. The SEM image shown in the lower right inset of Fig. 2b reveals a nanochannel network with arc-shaped nanochannels of about 40 nm high and about 150 nm wide. From this result, we confirm that the fabricated channels can scale down to several tens of nanometers.

A detailed AFM investigation in the left part of Fig. 2c (also indicated in the left part of Fig. 1c by a white dashed rectangle) reveals that a complicated surface distortion of the SiGe film develops near the etching front. From top to bottom, Fig. 2c shows that the periodicity of the wrinkling becomes larger but the morphology remains similar, which is known as the "self-similar folding" phenomenon and is generally found near the boundary between a fixed and free layer (e.g., the etching front in our experiments).[25, 26] The surface morphology of the SiGe film near the etching front was analyzed by an autocorrelation plot (right part of Fig. 2c), in which the maxima are marked as solid lines. The autocorrelation verifies that the periodicity of the wrinkles decreases towards the etching front, which has been explained previously using the competition between stretching and bending in a layer at a fixed boundary.[25, 26]

Customized linear nanochannel networks can be devised by mechanical scratching or lithography. We have also discovered that crystal defects in the film act as starting points for the underetching, in this way producing circular nanochannel networks as shown in Supplementary Fig. S2. Well-defined holes i.e. circular patterns with diameters ($d$) ranging from 20 nm to several tens of micrometers were fabricated by photo- and electron-beam lithography. We use these structures to understand the influence of the diameter ($D$) on the number of branch channels in the circular networks. Applying an identical etch-width ($L_e$= 2.5 µm) to all circular patterns, a statistical analysis of the number of branch channels $m$ as a function of the diameter $D$ ($D = d + 2L_e$) was carried out as shown in Fig. 3a (blue circles). The inset presents an optical microscopy image of an array of circular networks created by REBOLA.

It is safe to assume that the average periodicity of the branch channels in the circular network (which is calculated from the perimeter of the etching front divided by the number of branch channels) approaches $\lambda_0$, when the diameter tends to infinity and the circular becomes a linear network. In our study we find that the periodicity $\lambda$ of the branch channels as a function of $D$ can be phenomenologically fitted by the exponential decay function

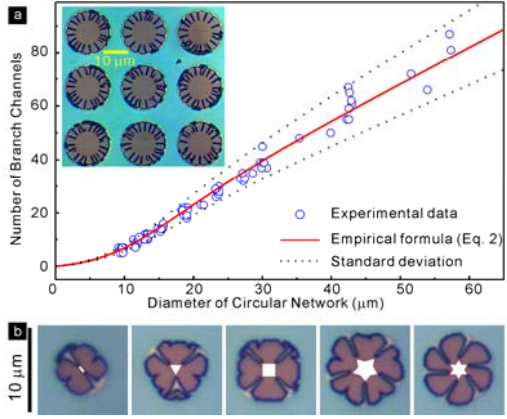

Figure 3: Size and shape effects on circular nanochannel networks. a, Number of branch channels versus diameter of circular network. Inset shows array of circular networks. b, Optical images of circular networks with 2-6 well-positioned branch channels. The circular networks evolve from a lithographically defined initial pattern (Defined pattern shapes form left to right:

$$\lambda = \lambda_0 + A \cdot \exp(-D/D_c). \qquad (1)$$

Here, $A$ is the amplitude of periodicity decaying and $D_c$ is the critical diameter. Using the relation $m\lambda = \pi D$ for circular networks, $m$ as a function of $D$ can be expressed as

$$m = \frac{\pi D}{\lambda_0 + A \cdot \exp(-D/D_c)}. \qquad (2)$$

For a good fit of our data in Fig. 3a (red line), we obtain $\lambda_0 = 2.30$ μm, $A = 12$ μm and $D_c = 6$ μm respectively. Here $\lambda_0$ corresponds to the average periodicity of a circular network with infinite diameter, and agrees well with the experimental value of 2.3 μm found for the corresponding linear network. The standard deviation of $\lambda_0$ from linear networks (17.1%) is introduced into our fit and explains the fluctuation of branch channel numbers in Fig. 3a. When $D$ drops below $D_c$, it is not possible to solve the number of branch channels with Eq. 2 because $D$ is comparable to or smaller than $2L_e$. To exert control over position and number of branch channels in circular networks with diameters of around $D_c$, well-defined initial patterns (indicated as white boxes in Fig. 3b) with special shapes were designed, such as short lines, triangles, squares, 5- and 6- fold stars. In these cases, the formation of the channels is initiated by the corners of each shape and well-positioned concentric branches are obtained in Fig. 3b.

One potential application of these nanochannel networks is their use in nanofluidics. Our proof of concept incorporates fluid transfer behaviour in a linear channel network and several individual wrinkles (see Supplementary Fig. S3), which have been pre-wetted with the applied fluid (ethylene glycol with 3.29E-04 molar rhodamine 6G). Fluid droplets of femto-litre (fl) volumes, created by a self-made drop generator, were introduced into this network as illustrated by a video microscopy frame in the left part of Fig. 4a (see entire video of fluid flow in the Supplementary Video). The right part of Fig. 4a is a schematic diagram of the droplet generator, which is based on the principle of thermal expansion of vapour bubbles in a glass capillary by heating the inserted thermal conductor metal wire (see details in Experimental section). The video microscopy documents the fluid flow through the linear nanochannel network, and the selected filling and emptying of a large wrinkle found in the lower left corner of Fig. 4a. (marked by a yellow dashed rectangle). Such a stand-alone wrinkle was previously referred to as a "blister".[2b] The filling is driven by corner flow through the linear network, which is caused by dissimilar chemical potentials of the fluid in the different channels.[27]

A time of 250 ms was needed for the fluid to flow from the position of the applied droplet to the large wrinkle. Selected chronological video frames, concentrating on the filling and emptying process of the large wrinkle, are given in Fig. 4b. The small branch channel of the upper linear network is situated at the very left side of the large wrinkle thus ensuring controlled fluid injection into the left part of the large wrinkle. The first frame at 0 ms was taken directly before filling of the large wrinkle started. From 0 to 350 ms, arc-like shapes of the fluid are observed in the large wrinkle. From 375 to 1575 ms, rectangular shapes are formed during fluid filling and subsequent emptying. In last frame (1600 ms), the arc shape appears again on the other side of the large wrinkle.

In order to understand the observed fluid shape transitions during the fluid filling and emptying, we consider a simplified geometry, which is sufficient to capture the main phenomena of our observations. The large wrinkle is modeled as a rectangular container with a volume $V_{cont}$, and the total surface energy $E_{surf}$ is calculated.[28] Two different fluid shapes, which are given by the Young-Laplace relation,[27,29] are considered (see insets of Fig. 4c). The change in surface energy as the channel fills and empties is shown in Fig. 4c. For shape 1 (upper right inset), the fluid is confined to one of the container walls while shape 2 (lower left inset) represents a square like fluid plug which extends from one wall to the other. In the very beginning (up to a relative volume of 0.04), the fluid is forced to the left edge and $E_{surf}$ increases due to the increase of surface area. Then, $E_{surf}$ gradually decreases as the fluid volume $V_f$ increases due to the hydrophilic nature of the fluid. In this volume regime, the fluid shape 2 experiences a lower surface energy than that of shape 1. However, the fluid cannot change from shape 1 to 2, since it has not reached the opposite wall, yet. Once the fluid reaches the opposite wall (which occurs for a normalized volume of 0.63), a rapid fluid shape transition occurs ("fluid shape transition I"). Continuous fluid supply into the channel keeps filling the channel. For emptying, we can use the same diagram as for filling. In this case, the fluid remains in shape 2 until a very small volume is occupied (normalized volume 0.10). For this volume, the surface energies cross and the fluid plug changes its shape to the one confined to the left wall ("fluid shape transition II") until the channel is completely emptied.

Figure 4d shows the measured fluid volume as a function of time. After complete filling at a constant flow rate of 23.2 fl/s from 375 to 600 ms, the large wrinkle empties with a constant slightly higher flow rate of 33.5 fl/s (from 1425 to 1600 ms). We note that the emptying process coincides with the almost complete evaporation of the generated droplet, which reverses the pressure conditions within the nanochannel network and drives the emptying of the large wrinkle.

Our studies suggests that an individual wrinkle might act as a good container for fluids of femto-litre volume, while the branch channels in the network represent an efficient fluid supplier or injector on the nanometer-scale. The experiments also demonstrate that small branch channels can deliberately be used to fill-up larger channels, which enables REBOLA to up-scale and in principle connect the nanochannel networks to the macroscopic world. We believe that the corner flow in such nanochannels may become increasingly important and interesting in nanofluidics because of its unique properties such as enhanced flow,[30] droplet generation within linear channels[31] and fluid mixing in networks.[32] REBOLA combines a bottom-up process – the wrinkling of a layer – with refined but standard main stream semiconductor technology (top-down process), and is thus expected to stimulate plenty of basic studies as well as various on-chip integrative applications.

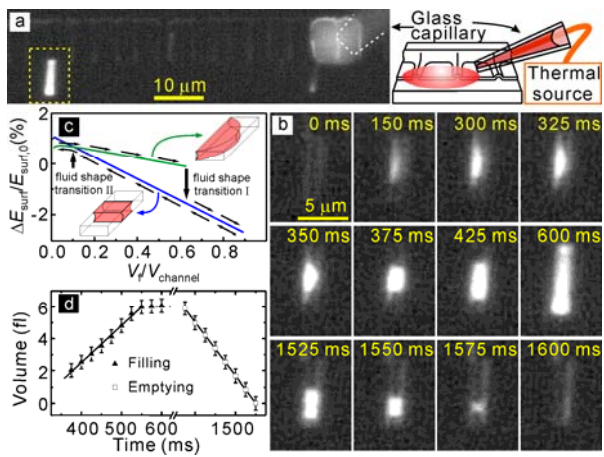

Figure 4: Femto-litre scale fluidics within linear nanochannel networks. a, Selected video frame (left) of liquid transport through a linear nanochannel network and a schematic diagram (right) of the experimental setup to generate a droplet. b, A series of selected video frames highlighting the transport behaviour of fluid in the large wrinkle. c, Surface energy as a function of fluid volume, in which the green line represents the arc-shaped fluid (upper right inset) and the blue line describes the rectangular-shaped fluid (lower left inset). d, Volume of fluid inside the wrinkle as a function of time.

Finally, we point out several advantages of REBOLA over other existing techniques: The size of the REBOLA channels is easily scalable by precisely engineering strain and thickness of the wrinkling layer as well as by the underetching length. Refined and expensive tools such as electron beam or extended ultra violet lithography are not necessarily required to fabricate nanometer sized wrinkles. The technique is compatible to advanced Si and CMOS technology. Furthermore, the top wrinkling layer is a single crystalline semiconductor layer and can easily be made optically or electronically active. This property might be beneficial to read out information stored in the fluid or to study interactions between the active semiconductor layer and the fluid.

In summary, we have demonstrated a technology (REBOLA) that exploits the deterministic wrinkling of a semiconductor layer to create well-defined and versatile nanochannel networks. In linear networks, the periodicity of branch channels as a function of etch-width was analyzed and compared with theoretical calculations. A self-similar folding phenomenon of wrinkles near a fixed boundary was revealed by autocorrelation analysis. The formation of branch channels within circular networks was studied on different length scales and was controlled by the size of the etched circular network and the shape of the initial pattern. To elucidate the usefulness of REBOLA, we exemplified nanofluidic transport as well as femto-litre filling and emptying of individual wrinkles on a standard semiconductor substrate, in which corner flow played an important role.

## Experimental

Fabrication of nanochannel networks: Ultra-thin silicon on insulator (SOI) wafers with a 27 nm top Si layer on 100 nm $SiO_2$ were loaded into an ultra-high vacuum, molecular beam epitaxy (UHV-MBE) system. A 40 nm layer of $Si_{1-x}Ge_x$ with uniform Ge composition of 8% was pseudomorphically grown on the ultra-thin SOI substrates at 350°C. After growth, the SiGe/SOI structures were oxidized in a tube furnace at 900°C for 100 minutes in oxygen ambient. The dry oxidation was followed by a 2 hours post-annealing step performed in nitrogen ambient. The grown $SiO_2$ layer was then removed by wet chemical etching, using a dilute HF acid solution. After defining the starting edge by mechanical scratching, or lithography followed by reactive ion etching (RIE), a 49% HF solution was used to selectively remove the buried $SiO_2$ layer resulting in the formation of nanochannel networks along the designed edges or patterns. Generally, we define long narrow lines by photo- or electron-beam lithography (width: 50 nm – several μm) for the single-sided linear branched channel networks, while two such lines with a tunable inter-distance were carefully designed to connect two single-sided branched channel networks to form a double-sided branched channel network. The etching distance is easily controlled by the duration of HF etching.

Fluidics setup: Observing the fluidic behaviour in the channel networks was accomplished by a Zeiss Axioskop upright microscope connected to an AxioCam MR for general optical colour images, and an AxioCam HSm for high-speed (60 – 198 frames/s) black-white videos. A rhodamine 6G specific fluorescence filter set was used to visualize the fluid transport behaviour. The glass capillaries (2 – 10 μm diameter tip) were manoeuvred using a PI NanoCube$^{TM}$ piezoelectric XYZ positioning stage mounted on a larger PI optics translation stage. With this combination a positioning accuracy of below 1 μm over a distance of 15 mm was achieved.

Solutions of rhodamine 6G in various solvents were created using a particulate lambdachrome rhodamine powder from Lambda Physik. Solvents used included ethanol and ethylene glycol of VLSI purity, and DI water. Concentrations of 5E-06 molar for ethanol, 3.29E-04 molar for ethylene glycol, and 4.14E-04 molar for water were used.

The femto-litre generator as shown in the right part of Fig. 4a is self-made and is based on the principle of thermal expansion of bubbles in the glass capillary as shown in the right part of Fig. 4a. Metal wire (polymer insulated copper 0.4 mm in diameter) was used as a thermal conductor connected to the thermal source (thermal iron) and immersed in the fluid (ethylene glycol with 3.29E-04 molar dye) in the glass capillary to heat the vapour bubbles. Heating the metal wire for a short time (several seconds) caused bubbles in the capillary to expand creating a pressure inside the capillary. This generates a fluid droplet with femto-litre volume out of the capillary tip.

[1] A. K. Harris, P. Wild, and D. Stopak, *Science* **1980**, *208*, 177.
[2] a) T.-Y. Zhang, X. Zhang, and Y. Zohar, *Journal of Micromechanics and Microengineering* **1998**, *8*, 243. b) X. Zhang, Y. Zohar, and T.-Y. Zhang, *Micro-Electro-Mechanical Systems (MEMS)* **1998**, DSC-Vol. 66, 379.
[3] N. Bowden, S. Brittain, A. G. Evans, J. W. Hutchinson, and G. M. Whitesides, *Nature* **1998**, *393*, 146.
[4] O. G. Schmidt, N. Schmarje, C. Deneke, C. Muller, and N. Y. Jin-Phillipp, *Advanced Materials* **2001**, *13*, 756.
[5] E. Cerda, K. Ravi-Chandar, and L. Mahadevan, *Nature* **2002**, *419*, 579.
[6] V. Ya. Prinz, *Microelectronics Engineering* **2003**, *69*, 466.
[7] E. Cerda and L. Mahadevan, *Physical Review Letters* **2003**, *90*, 074302.
[8] P. J. Yoo and H. H. Lee, *Physical Review Letters* **2003**, *91*, 154502.
[9] C. M. Stafford, C. Harrison, K. L. Beers, A. Karim, E. J. Amis, M. R. Vanlandingham, H.-C. Kim, W. Volksen, R. D. Miller, and E. E. Simonyi, *Nature Materials* **2004**, *3*, 545.

[10] K. Efimenko, M. Rackaitis, E. Manias, A. Vaziri, L. Mahadevan, and J. Genzer, *Nature Materials* **2005**, *4*, 293.
[11] D.-Y. Khang, H. Jiang, Y. Huang, and J. A. Rogers, *Science* **2006**, *311*, 208.
[12] T. M. Squires and S. R. Quake, *Reviews of Modern Physics* **2005**, *77*, 977.
[13] J. C. T. Eijkel and A. v. d. Berg, *Microfluidics and Nanofluidics* **2005**, *1*, 249.
[14] P. J. A. Kenis and A. D. Stroock, *MRS Bulletin* **2006**, *31*, 87.
[15] M. A. Northrup, *Nature Materials* **2004**, *3*, 282.
[16] J. W. Hong, V. Studer, G. Hang, W. F. Anderson, and S. R. Quake, *Nature Biotechnology* **2004**, *22*, 435.
[17] P. S. Dittrich and A. Manz, *Nature Reviews Drug Discovery* **2006**, *5*, 210.
[18] O. G. Schmidt and K. Eberl, *Nature* **2001**, *410*, 168.
[19] O. G. Schmidt, Y. Mei, D. Thurmer, and F. Cavallo, *European Patent*, **Jan 2006** (filed).
[20] A. I. Fedorchenko, A.-B. Wang, V. I. Mashanov, and H.-H. Cheng, *Journal of Mechanics* **2005**, *21*, 131.
[21] J. C. Tsang, P. M. Mooney, F. Dacol, and J. O. Chu, *Journal of Applied Physics* **1994**, *75*, 8098.
[22] $x$ is the Ge composition, $\Delta_{Si}$ the Raman shift normalized by $x = 1$ (i.e. assuming 100% composition of Ge in the SiGe layer), and $\Sigma$ is the normalized strain of the measured film using the mismatch strain 0.0417 (i.e. the strain between pure Ge and pure Si making $\Sigma = 1$ for pure Ge grown epitaxially on Si (001)).
[23] A more detailed calculation will be published elsewhere.
[24] G. M. Cohen, P. M. Mooney, V. K. Paruchuri, and H. J. Hovel, *Applied Physics Letters* **2005**, *86*, 251902.
[25] S. Conti, A. DeSimone, and S. Müller, *Computer Methods in Applied Mechanics and Engineering* **2005**, *194*, 2534.
[26] E. Cerda, L. Mahadevan, and J. M. Pasini, *Proceedings of the National Academy of Science* **2006**, *101*, 1806.
[27] J. C. T. Eijkel and A. v. d. Berg, *Lab on a Chip* **2005**, *5*, 1202.
[28] $E_{\text{surf}}$ is calculated by $E_{surf} = \gamma_{sf} A_{sf} + \gamma_{fv} A_{fv} + \gamma_{sv} A_{sv}$, where, *s*, *f*, and *v* denote solid, fluid, and vapor phase, respectively; $\gamma_{ij}$ is the surface tension between phase *i* and *j*; $A_{ij}$ is the surface area between phase *i* and *j*. The surface energy without any fluid is $E_{surf,0} = \gamma_{sv} A_{sv}$, while the relative surface energy $\Delta E_{surf}/E_{surf,0} = (E_{surf} - E_{surf,0})/E_{surf,0}$ is considered. The simplified channel geometry is 1.7 µm wide, 470 nm high, and 7.5 µm long.
[29] N. R. Tas, P. Mela, T. Kramer, J. W. Berenschot, and A. v. d. Berg, *Nano Letters* **2003**, *3*, 1537.
[30] J. C. T. Eijkel, B. Dan, H. W. Reemeijer, D. C. Hermes, J. G. Bomer, and A. v. d. Berg, *Physical Review Letters* **2005**, *95*, 256107.
[31] P. Garstecki, M. J. Fuerstman, H. A. Stone, and G. M. Whitesides, *Lab on a Chip* **2006**, *6*, 437.
[32] D. Belder, M. Ludwig, L. W. Wang, and M. T. Reetz, *Angewandte Chemie International Edition* **2006**, *45*, 2463.